\documentclass[epj]{svjour}
\pdfoutput=1

\usepackage{graphics}
\usepackage[title]{appendix}
\usepackage{amssymb,amsmath,rotate,color}
\usepackage{bm}
\usepackage{soul}

\usepackage{graphicx}
\usepackage{dcolumn}
\usepackage{bm}
\usepackage{bbm}
\begin{document}
\title{Modified spin-orbit couplings in uniaxially strained graphene}
\author{H. Rezaei\inst{1,2}  \and A. Phirouznia \inst{1,2,3} 
}                     
%
%
\institute{Department of Physics, Azarbaijan Shahid Madani University, 53714-161, Tabriz, Iran \and Condensed Matter
 Computational Research Lab. Azarbaijan Shahid Madani University 53714-161, Tabriz, Iran  \and Computational Nanomaterials Research Group (CNRG) 
Azarbaijan Shahid Madani University 53714-161, Tabriz, Iran}
\date{Received: date / Revised version: date}
%
\abstract{Intrinsic and Rashba spin-orbit interactions in strained graphene is studied within the tight-binding (TB) approach. Dependence of Slater-Koster (SK) parameters of graphene on strain are extracted by fitting the \emph{ab initio} band structure to the TB results. A generalized low-energy effective Hamiltonian in the presence of spin-orbit couplings is proposed for strained graphene subjected to an external perpendicular electric field. Dependence of the modified Rashba strength and other parameters of effective Hamiltonian on the strain and electric field are calculated. In order to analyze the influence of the applied strain on the electronic properties of the graphene, one must take into account the lattice deformation, modifications of the hopping amplitudes and shift of the Dirac points. We find that using the strain it is possible to control the strength of Rashba and intrinsic spin-orbit couplings as well as energy gap at the shifted Dirac points. Meanwhile, the strain slightly modifies the topology of low-energy dispersion around the Dirac points. We describe the SOCs induced energy splitting as a function of strain. 
\PACS{
      {73.22.-f}{Electronic structure of nanoscale materials and related systems}   \and
      {71.70.Ej}{Spin-orbit coupling, Zeeman and Stark splitting, Jahn-Teller effect}
      {71.70.Fk}{Strain-induced splitting}
       } 
} 
\maketitle
\section{Introduction}
Graphene has been the subject of intense investigations due to its outstanding electronic and mechanical properties \cite{Castro}. The most notable electronic property of graphene is its linear gapless energy dispersion around the so called Dirac points ($K$ and $K'$ points) at low energy regime. However, it has been shown that spin-orbit couplings (SOCs) in graphene, slightly change the gapless linear band structure of graphene and can open up an energy gap at the Dirac points \cite{Macdonald}. The magnitude of the intrinsic spin-orbit induced gap has been the subject of discussions by researchers of this filed \cite{Macdonald,firstPrncpl9,firsPrncpl49,Fabian,FabianTopolgy50,kane7}. Applying an external electric field perpendicular to the graphene sheet causes Rashba type SOC which can be regarded as external SOC. Although the spin-orbit interaction in the graphene is weak \cite{nanotube_10}, it plays an important role in the half integer quantum Hall effect, spintronics and spin dependent properties \cite{nature_spintronic_80,Macdonald,kane7,Obtic_86,SOC_88}. 

Another attracting field in the graphene research is the strain induced effects on the electronic properties. Electronic structure of strained graphene has been studied by several authors \cite{engineering_39,Strain_tensor_26,optical_31,guinea_strain_36,zhan2012engineering_69}. Strain modifies the electronic properties of graphene. For instance, strain changes the position of the Dirac points and hopping amplitude \cite{mexic}. Investigating the SOCs in strained graphene is the subject of the present study.

Intrinsic spin-orbit coupling-induced band gap of the graphene under strain has been studied by B. Gong \emph{et al.} using \emph{ab initio} calculations and tight-binding (TB) method \cite{BAIHUA}. They have used Harrison's expression to formulate the dependence of the hopping parameters on strain, where they showed that the energy gap has a monotonic increasing dependence on the strain. Meanwhile, within the most of research in this field, the hopping amplitude is assumed to be modified as a result of the change in atomic distances and the lattice deformation which determines the neighboring orbitals orientation is not considered. However, in the case of uniaxial strains, orbitals reorientation must be taken into account. An effective Hamiltonian of intrinsic SOC for strained graphene has been extracted by employing the method of invariants and \emph{ab initio} calculations in the vicinity of Dirac points at very small strains \cite{MoS_29}. In another work by G. S. Diniz \emph{et al.} manipulation of the quantum anomalous Hall effect in graphene as a result of the applied strain has been studied \cite{engineering_39}.
   
In this article we have studied the intrinsic and Rashba spin-orbit interactions in the strained graphene using a TB model within the subspace of s and p orbitals at low-energy regime\cite{Macdonald,sp_82,low_11,nanotube_10}. It has been shown that the strain accountably changes the strength and functionality of the spin-orbit interactions. Present study has been limited to in-plane uniaxial strains in the range of -20\% to 20\% which was assumed to be applied in either zigzag or armchair direction. In order to study the electronic properties of strained graphene both the lattice deformation and dependence of the hopping parameters on strain must be taken into account. In addition as shown in the next sections unlike the homogeneous strain, uniaxial strain could change the neighboring orbitals orientation which can change the hopping and therefore hopping related parameters such as Rashba coupling strength. At the first step one has to parameterize the dependence of the hopping amplitudes on the applied strain. Then, TB Hamiltonian in the presence of strain and SOCs can be expressed as an effective low energy Hamiltonian at shifted Dirac points. L\"{o}wdin method have been employed to extract the effective low-energy Hamiltonian at the shifted Dirac points \cite{firstPrncpl9,Lowdin1}.
Due to the similar nature of the Dirac materials it is expected that the present approach could be extended to study of other honeycomb structures such as stanene, germane and silicene.  \cite{YAKOVKIN20171,Dirac_material,low_11}.     

\section{Graphene in the presence of strain:}

\subsection{Tight-Binding model of graphene}
\begin{figure}
	\resizebox{0.5\textwidth}{!}{
			\includegraphics{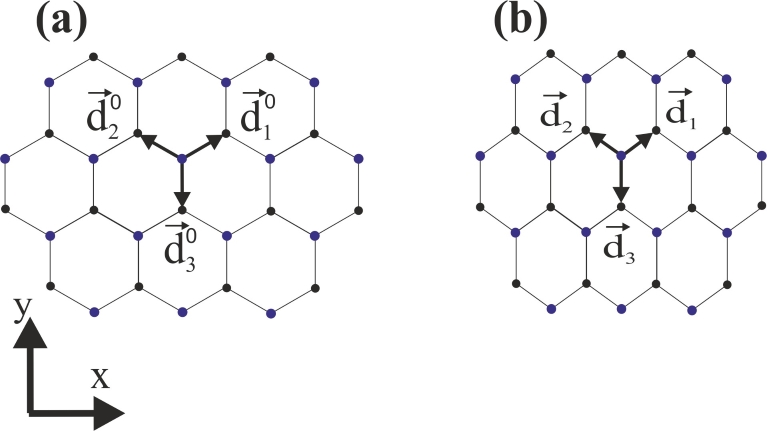}
			}
	\caption{(Color online) (a) Unstrained Graphene lattice (b) Zigzag strained graphene lattice. $\vec{d_{i}^{0}}$ and $\vec{d_{i}}$ refer to the i'th nearest-neighbor vectors in the unstrained and strained graphene respectively. \label{Fig_vector}}
\end{figure}

Two-center Slater-Koster (SK) nearest-neighbor TB method \cite{slater23} has been employed with s and p orbitals to calculate the electronic structure of strained graphene at low-energy scheme in which the intrinsic and external spin-orbit couplings have been considered. In the unstrained graphene nearest-neighbor atoms are connected by three vectors

\begin{equation}
\label{vec}
\vec{d_1^{0}}=\frac{a_0}{2}(\sqrt{3},1),   \vec{d_2^{0}}=\frac{a_0}{2}(-\sqrt{3},1),  \vec{d_3^{0}}=a_0(0,-1),
\end{equation}
where $a_0= 1.42 {\AA}$ is the carbon-carbon distance in unstrained graphene. Strained and unstrained graphene lattice and nearest-neighbor vectors have been shown in Fig.\ref{Fig_vector}. We choose the x axis in a way that is parallel to the zigzag direction of honeycomb lattice. In the nearest-neighbor TB model for uniaxially strained graphene in the absence of SOCs, Hamiltonian matrix elements are given by

\begin{equation}
{\label{eq_TB_H}
	\emph{H}^{AB}_{l,m}(\vec{k}) = {\emph{H}^{*}}^{BA}_{m,l}(\vec{k}) = \sum _{i=1}^{3}{t_{l,m}(\vec{d_{i}})e^{i \vec{k}.\vec{d_{i}}}}},
\end{equation}

\begin{equation}
{\emph{H}^{AA}_{l,m}(\vec{k})=\emph{H}^{BB}_{l,m}(\vec{k})=E_{l}\delta_{l,m}}.
\end{equation}

Here, $A$ and $B$ refer to different sublattices, k is the wave vector, $\vec{d_{i}}$ represents the nearest-neighbor position vector in the strained lattice as shown in Fig. \ref{Fig_vector}, $t_{l,m}(\vec{d_{i}})$ are the hopping matrix elements between $l$ and $m$ orbitals in the nearest $i$'th neighbor site and $E_{l}$ is the energy of $l$'th orbital. We should notice that in the strained graphene $t_{l,m}(\vec{d_{i}})$ and $\vec{d_{i}}$ depend on strain tensor $\bm{\epsilon}$. The relation between hopping matrix elements and SK parameters $V_{pp\pi}, V_{pp\sigma}, V_{sp\sigma}$ and $V_{ss\sigma}$ are listed in Table \ref{tab_Slater}. Since in general orbitals in the neighboring atoms are not orthogonal, we need to include non-zero overlap parameters, $S_{lm}$, in the computational approach. However, at the Dirac points it is possible to neglect the overlap parameters for simplicity \cite{Fabian,Macdonald}.

 Without taking into account the spin degree of freedom, TB Hamiltonian of graphene in the absence of SOC (by considering one $s$ and three $p$ orbitals of the outer shell of carbon atoms) can be represented by $8\times8$ block-diagonal matrix containing a $2\times2$ $\pi$ block and $6\times6$ $\sigma$ block which result in $\pi$ and $\sigma$ bands respectively. $s$, $p_{x}$ and $p_{y}$ orbitals results in $\sigma$ bands while the other out-of-plan $p_{z}$ orbitals create $\pi$ bands. In the vicinity of the Dirac points the electronic properties can be described by $\pi$ block that causes Dirac-type Hamiltonian. Spectrum of this Hamiltonian is gapless and linear which results in Dirac cones \cite{Castro}. Following many other previous articles \cite{Macdonald,sp_82,low_11,firstPrncpl9,nanotube_10} in the present study the $sp$ based TB model has been considered to study the SOCs in strained graphene. However, some authors propose using $d$ orbitals as well as $s$ and $p$ orbitals to analyze SOCs in graphene \cite{Fabian,FabianTopolgy50}. As they discussed using $d$ orbitals results in intrinsic gap of the order of 25 $\mu eV$ while in $sp$ model intrinsic gap of unstrained graphene is about 1 $\mu eV$ as calculated in the present paper for unstrained graphene. Meanwhile, they have reported that the Rashba coupling is dominated by the $s$ and $p$ orbitals, namely, $\pi-\sigma$ coupling \cite{Fabian}.

\begin{table}
	\caption{Hopping matrix elements between s and p orbitals in terms of two-center SK parameters $V_{\mu}$ along the unit vector $(n_{x},n_{y},n_{z})$\label{tab_Slater}}
	\begin{tabular}{lcll}
		\hline
		\hline
		\\
		$t_{s}$     ~~& $s$                &~~~~~~~~~~ $t_{x,x}$   ~~~~& $n_{x}^{2}V_{pp\sigma}+(1-n_{x}^{2})V_{pp\pi}$ \\
		$t_{p}$     ~~& $p$                &~~~~~~~~~~ $t_{y,y}$   ~~~~& $n_{y}^{2}V_{pp\sigma}+(1-n_{y}^{2})V_{pp\pi}$ \\
		$t_{s,s}$   ~~& $V_{ss\sigma}$     &~~~~~~~~~~ $t_{z,z}$   ~~~~& $n_{z}^{2}V_{pp\sigma}+(1-n_{z}^{2})V_{pp\pi}$ \\
		$t_{s,x}$   ~~& $n_{x}V_{sp\sigma}$&~~~~~~~~~~ $ t_{x,y}$  ~~~~& $n_{x}n_{y}(V_{pp\sigma}-V_{pp\pi})$   \\
		$t_{s,y}$   ~~& $n_{y}V_{sp\sigma}$&~~~~~~~~~~ $ t_{x,z}$  ~~~~& $n_{x}n_{z}(V_{pp\sigma}-V_{pp\pi})$   \\
		$t_{s,z}$   ~~& $n_{z}V_{sp\sigma}$&~~~~~~~~~~ $ t_{y,z}$  ~~~~& $n_{y}n_{z}(V_{pp\sigma}-V_{pp\pi})$   \\
		\\
		\hline
		\hline
	\end{tabular}
\end{table}

For a two-dimensional structure strain tensor is given as
\begin{equation}
\bm{\epsilon}=\left(
           \begin{array}{cc}
             \epsilon_{xx} & \epsilon_{xy} \\
             \epsilon_{xy} & \epsilon_{yy} \\
           \end{array}
         \right),
\end{equation}
in which the uniaxial strain tensor can be written as follows \cite{Strain_tensor_26}
\begin{equation}
\bm{\epsilon}=\epsilon\left(
           \begin{array}{cc}
             cos^{2}\theta-\nu sin^{2}\theta & (1+\nu)cos \theta  sin \theta \\
             (1+\nu)cos \theta  sin \theta & sin^{2}\theta-\nu cos^{2}\theta \\
           \end{array}
         \right),
\end{equation}
where $\theta$ is the angle between the direction of strain and x axis and $\nu\approx 0.14$ is the Poisson's ratio.\cite{Poisson,optical_31}. The relation between the displacement vectors of the nearest neighbor atoms in the unstrained and strained graphene can be written as
\begin{equation}
\vec{d}_{i}=(\textbf{I}+\bm{\epsilon)}.\vec{d}^{0}_{i},
\end{equation}
where $\textbf{I}$ is the $2\times2$ identity matrix.

It should be noted that applying a strain in some situations (for example uniaxial strains larger than 20\% in a given specific direction) can open up a gap in the low-energy spectrum of graphene\cite{exponen27,zhan2012engineering_69}. However it is not the case for strains $\epsilon\leq 0.2 $ as in our work.
\subsection{Slater-Koster Parameters }
By fitting the numerical results of the TB method to the \emph{ab initio} calculations we can deduce the hopping and overlap parameters of graphene. The ABINIT package has been employed for non-relativistic \emph{ab initio} calculations of strained and unstrained graphene \cite{Abinit1,Abinit2}, where the band energy of the single layer graphene has been obtained for different strains in order to calculate SK parameters and dependence of these parameters on strain. A $32 \times 32 \times 1$ Monkhorst-Pack \cite{mon-pak1976} mesh grid has been employed in the first Brillouin zone sampling for discretization of the Kohn-Sham equations. A vacuum space of 30 Bohr is placed to avoid atomic orbital overlap between the given monolayer and its periodic images. Meanwhile, local density approximation (LDA) has been considered for exchange-correction energy functional. This is necessary for eliminating the interaction between the periodic layers which are generated in the plane-wave based solutions of the Kohn-Sham equations. Maximal kinetic energy cut-off is $50$ Hartree. It should be noted that the given parameters result in proper convergence of the total energy. 

\begin{figure}
\includegraphics[width=0.9\linewidth]{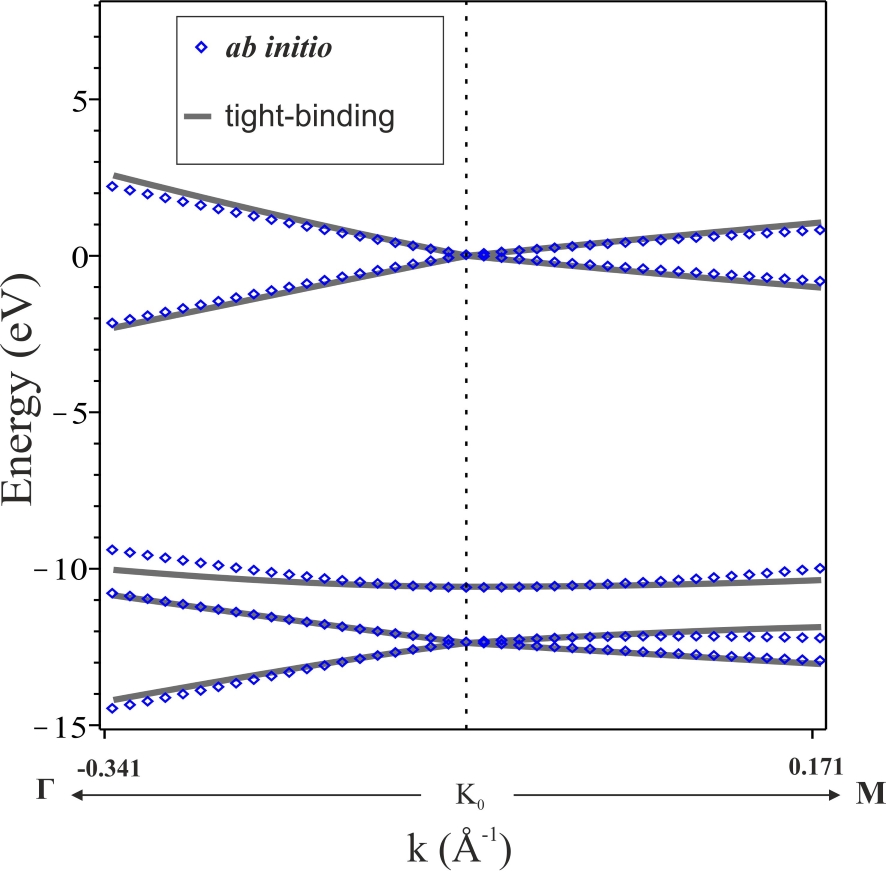}
\caption{(Color online) Unstrained graphene band structure using \emph{ab initio} calculations (diamonds) and TB method (solid lines). Calculations have been carried out in the vicinity of the Dirac point along the $\Gamma$-K-M direction. \label{figBandStructure}}
\end{figure}

We have focused on low-energy effective Hamiltonian around the Dirac point, therefore, the fitting calculations have been performed in the vicinity of Dirac points. Fig.\ref{figBandStructure} presents \emph{ab initio} and TB band structure of unstrained graphene in the range of points in the $k$-space in which the TB energy bands has been given using the optimized parameters from the first-principle calculations. As shown in this figure the band energies in the path $\Gamma$-$K$-$M$ containing the $K$ point (Dirac point) are in good agreement with the \emph{ab initio} calculations. However, it should be noted that some of the upper conduction bands given by \emph{ab initio} approaches cannot be fitted to TB results \cite{PhysRev70}. The results of the present numerical calculation for SK parameters have been compared with two other reports Refs. \cite{Fabian,Macdonald} as shown in Table \ref{tabHopping}. As can be seen in this table, our results are at the same range of the other reports and it is clear that the TB parameters and band structure that have been obtained in the present study show a better coincidence with the result of the Ref. \cite{Fabian}. The slight difference may be originated from the fact that the fitting range in the current work lies around the Dirac point instead of all high-symmetry points of the Brillouin zone. 

\begin{table*}
\caption{Slater-Koster parameters of unstrained graphene. Comparison between our results and other references. (a) Present results (b): Ref.\cite{Fabian} (c): Ref.\cite{c60-SK-param-53}. $\Delta=\varepsilon_{p}-\varepsilon_{s}.$ Present values are calculated by fitting the TB results to the \emph{ab initio} calculations in the vicinity of the Dirac point \label{tabHopping}}
\begin{tabular}{llllll}
  \hline
  \hline
 \\
  Parameter   ~~~     & $V_{ss\sigma}$    & $V_{sp\sigma}$    & $V_{pp\sigma}$ & $V_{pp\pi}$ & $\Delta$  \\
  $Energy(eV)^{a}~~~$ &-5.71~~~           & 5.42~~~           & 6.20~~~        & -3.07~~~    & -8.37 \\
  $Energy(eV)^{b}~~~$ & -5.729~~~         & 5.618~~~          & 6.05~~~        & -3.07~~~    & -8.37  \\
  $Energy(eV)^{c}~~~$ & -6.769~~~         & 5.58~~~           & 5.037~~~       & -3.033~~~   & -8.868  \\
\\
  \hline

\\
    Parameter ~~~& $S_{ss\sigma}$ &  $S_{sp\sigma}$ & $S_{pp\sigma}$ & $S_{pp\pi}$& ~~~~~~~ \\
  $Value^{a}$ ~~~& 0.10~~~        & -0.170~~~       & -0.140~~~      & 0.07~~~    & ~~~~~~~ \\
  $Value^{b}$ ~~~& 0.102~~~       & -0.171~~~       & -0.377~~~      & 0.07~~~    & ~~~~~~~ \\
  $Value^{c}$ ~~~& 0.212~~~       & -0.102~~~       & -0.146~~~      & 0.129~~~   & ~~~~~~~ \\
  \hline
  \hline
\\
\end{tabular}
\end{table*}

\subsection{Dependence of hopping parameters on strain}
\begin{figure}
\includegraphics[width=0.85\linewidth]{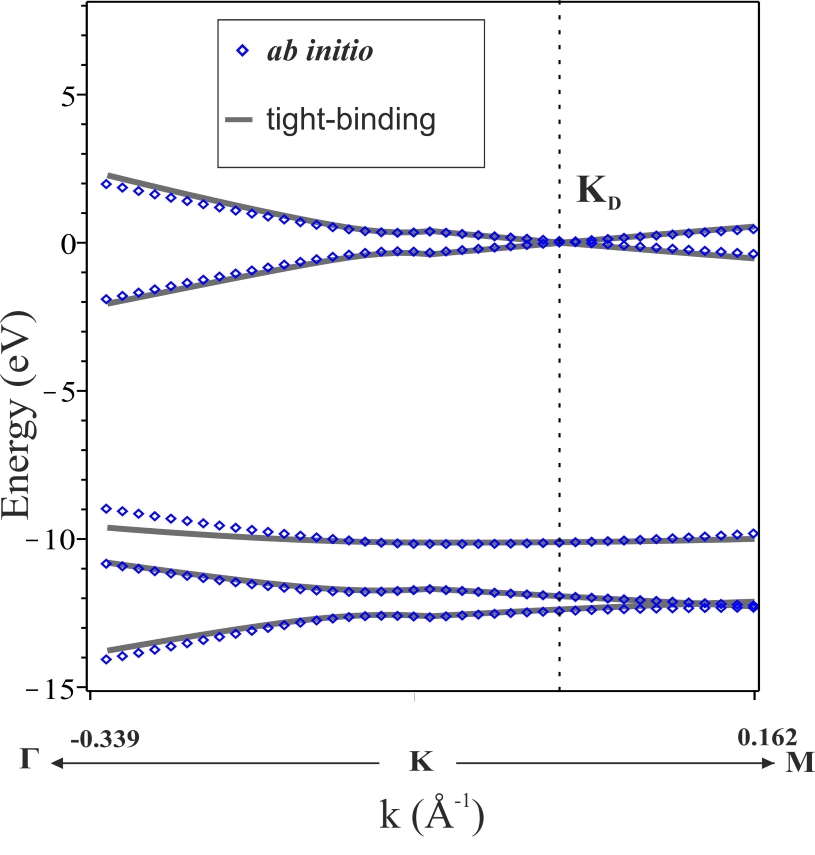}
\caption{(Color online) Graphene band structure under zigzag strain of 5\% using \emph{ab initio} calculations (diamonds) and TB method (solid lines). Calculations have been carried out in the vicinity of the shifted Dirac point ($K_{D}$) along the $\Gamma$-K-M direction in the  distorted reciprocal space. Note that, the point $K$ is the corner of the strained Brillouin zone and not the Dirac point.
\label{badstruct_strain}}
\end{figure}

SK hopping parameters in the strained graphene depend on the magnitude and direction of the strain. The variation of the SK hopping parameters with the strain can be modeled by an exponential relation \cite{exponen27,engineering_39,Strain_tensor_26} such as

\begin{equation}
V_{\mu}(d)=V_{\mu}^{0} exp(-\beta_{\mu} (\frac{|\vec {d}|}{a_{0}}-1)),
\end{equation}
where $V_{\mu}$ and $V_{\mu}^{0}$ are the SK parameters of type $\mu$ in the strained and unstrained graphene respectively, $\beta_{\mu}$ is a parameter that characterizes the influence of the strain on the SK parameters and must be determined for each SK parameter separately. Parameter $\beta$ for $\pi$ band have been calculated by some authors and can be estimated as $\beta_{pp\pi} \sim 3$ \cite{mexic,exponen27,Strain_tensor_26}. In the current work numerical values of the parameter $\beta$  for all SK parameters (hopping amplitudes and overlap parameters) that needed in the present approach have also been estimated. As mentioned within the current approach it is possible to extract other  $\beta$ parameters by fitting the \emph{ab initio} and TB results for strained graphene. Results have been shown in Table \ref{tabHopping} and Fig.\ref{badstruct_strain}. The value for $\beta_{pp\pi}$ are in good agreement with the values of Refs.\cite{Strain_tensor_26,exponen27} and \cite{mexic}. Some authors have studied the relation between the other SK parameters and the modified atomic separation \cite{parameterization_78,konschuhThesis6}. It is worthwhile to compare those results with the numerical results of the present work. 

\begin{table}
\caption{$\beta$ parameters calculated by fitting the TB results to the \emph{ab initio} calculations. $\beta$ and $\beta '$ refer to hopping and overlap parameters respectively  \label{tab:Beta}}
\begin{tabular}{lcccc}
  \hline
  \hline
\\
  Parameter ~~~~ & $\beta_{ss\sigma}~~~~$ & $\beta_{sp\sigma}~~~~$ & $\beta_{pp\sigma}~~~~$ & $\beta_{pp\pi}$  \\
\\
  Value ~~~~  & 3.17~~~~         & 1.82~~~~         & 1.47~~~~           & 3.104         \\
\\
  \hline
\\

  Parameter   ~~~~& $\beta^{'}_{ss\sigma}~~~~$ &  $\beta^{'}_{sp\sigma}~~~~$ & $\beta^{'}_{pp\sigma}~~~~$ & $\beta^{'}_{pp\pi}$ \\
\\
  Value ~~~~&  2.72~~~~             & 1.28~~~~               & 0.77~~~~              & 2.11~~~~           \\
\\
  \hline
  \hline
\end{tabular}
\end{table}

\section{Spin-orbit couplings in the graphene}
Intra-atomic spin-orbit interaction term in the graphene structure can be written as \cite{low_11} 
\begin{equation}
\emph{H}_{SO}=\xi \vec{L}. \vec{s},
\end{equation}
where $\xi\sim 6meV $ is the spin-orbit coupling constant among the p orbitals \cite{Macdonald}, $\vec{L}$ is the angular momentum operator and $\vec{s}$ stands for the spin of electron. Considering $H_{SO}^{AA}=H_{SO}^{BB}$ and $H_{SO}^{AB}=0$, it is possible to deduce all on-site SOC Hamiltonian matrix elements.

In the presence of an external electric field which applied perpendicularly to the graphene sheet, the contribution of the Stark effect can be considered as
\begin{equation}
H_{Stark}=-e E \hat{z},
\end{equation}
where $e$ is the electron charge, $E$ is the strength of the electric field and $\hat{z}$ is the position operator along the $z$ axis. The only non-zero elements of the Stark Hamiltonian matrix are on-site coupling between $s$ and $p_{z}$ orbitals which can be written as $\langle p_{z}|H_{Stark}|s\rangle =eE z_{sp}$, where $z_{sp}$ is the electric dipole transition between $s$ and $p$ orbitals \cite{Fabian}. This electric field which can be originated from a gate voltage, breaks the inversion symmetry in the graphene plane. Now we can write down the $H_{Stark}+H_{SO}$ matrix in the subspace of the following basis:\\
$\{ p_{z}\uparrow,p_{z}\downarrow,s\uparrow,s\downarrow,p_{x}\uparrow,p_{x}\downarrow,p_{y}\uparrow,p_{y}\downarrow\}$\\
as given by
\begin{equation}
H_{Stark}+H_{SO}=\frac{1}{2}\xi\left(
\begin{array}{cccccccc}

0         & 0         & E_{S}     & 0         & ~0~  &    -1 &   0 &   i  \\
0         & 0         & 0         & E_{S}     & ~1~  &     0 &   i &   0  \\
E_{S}     & 0         & 0         & 0         & ~0~  &     0 &   0 &   0  \\
0         & E_{S}     & 0         & 0         & ~0~  &     0 &   0 &   0  \\
0         & 1         & 0         & 0         & ~0~  &     0 &  -i &   0  \\
-1         & 0         & 0         & 0         & ~0~  &     0 &   0 &   i  \\
0         &-i         & 0         & 0         & ~i~  &     0 &   0 &   0  \\
-i         & 0         & 0         & 0         & ~0~  &    -i &   0 &   0  \\
\end{array}
\right),
\end{equation}
In which $E_{S}=\frac{2}{\xi}eE z_{sp}$.
\section{Spin-orbit couplings in strained graphene}
\subsection{Generalized effective Hamiltonian}
In this sections the spin-orbit couplings in the strained graphene have been studied at low-energy regime. Because the intrinsic SOC in graphene is weak (that can be inferred from the small SOC constant $\xi$ in the graphene), it can be regarded as a perturbation in the TB Hamiltonian \cite{nanotube_10}. The total Hamiltonian of the system reads 
\begin{equation}
H=H_{0}+H_{SO}+H_{Stark},
\end{equation}
$H_{0}$ stands for the TB Hamiltonian of strained graphene without SOCs. This Hamiltonian can be divided into four blocks
\begin{equation}
H_=
\left(
\begin{array}{cc}
H_{\pi} & T \\
T^{\dag} & H_{\sigma} \\
\end{array}
\right).
\end{equation}

$H_{\pi}$ denotes the $\pi$ band block, $H_{\sigma}$ describes the $\sigma$ band and these two blocks coupled by $T$ block which depends basically on the SOC. The representation basis of $H_{\pi}$ is $\{Ap_z\uparrow,~ A p_{z}\downarrow,~ B p_{z}\uparrow,~B p_{z}\downarrow \}$, while the basis set of $H_{\sigma}$ is $\{A s\uparrow, A s\downarrow, A p_{x}\uparrow, A p_{x}\downarrow,A p_{y}\uparrow, A p_{y}\downarrow,B s\uparrow, B s\downarrow, B p_{x}\uparrow, B p_{x}\downarrow,B p_{y}\uparrow, B p_{y}\downarrow\} $ the directed atomic orbitals. By performing the L\"{o}wdin partitioning on the Hamiltonian, one could obtain the $4\times4$ effective strain modified SOC Hamiltonian at the shifted $K(K')$ points.

In the strained graphene the $4\times4$ effective SOC Hamiltonian at the Dirac points can be calculated which finally could be proposed at the following form:
\begin{eqnarray}
\label{Eff}
H_{eff}&=&(\delta_{SO}-\lambda_{SO})\mathbbm{1}_{\sigma s}+  \lambda_{SO}\tau\sigma_{z}s_{z}+  (\lambda_{R}-\delta_{R})\tau \sigma_{x}s_{y}\nonumber\\
&&-(\lambda_{R}+\delta_{R})\sigma_{y}s_{x}-\delta_{0}\tau \mathbbm{1}_{\sigma} s_{y},
\end{eqnarray}
in which $\tau$ is the valley index, $\mathbbm{1}_{\sigma s}\equiv \mathbbm{1}_{\sigma} \otimes~ \mathbbm{1}_{s} $ is the $4\times4$ identity matrix, $\mathbbm{1}_{\sigma}$ and $\mathbbm{1}_{s}$ are the $2\times2$ identity matrices in pseudospin and spin spaces respectively. $\lambda_{SO}$, $\lambda_{R}$, $\delta_{SO}$, $\delta_{R}$ and $\delta_{0}$ are strain dependent parameters of effective SOC Hamiltonian at the Dirac points. A given term like $\sigma_{\alpha}s_{\beta}$ is a short notation of $\sigma_{\alpha}\otimes s_{\beta}$. This effective Hamiltonian can be regarded as a generalized form of the Hamiltonian represented by Ref.\cite{Macdonald}. In the absence of strain $\delta_{SO}$, $\delta_{R}$ and $\delta_{0}$ parameters vanish identically. The magnitude of $\lambda_{SO}$ and $\lambda_{R}$ in the strained graphene is different form their unstrained values ($\lambda^{(0)}_{SO}$ and $\lambda^{(0)}_{R}$). In the basis set of \{$Ap_z\uparrow,~ A p_{z}\downarrow,~ B p_{z}\uparrow,~B p_{z}\downarrow $\} the matrix elements of this effective Hamiltonian (for $\tau=-1$) can be represented as 

\begin{equation}
H_{eff}=
\left(
\begin{array}{cccc}
\delta_{SO}-2\lambda_{SO}~~ & -i\delta_{0}~~~ & 0             & 2i\lambda_{R} \\
i\delta_{0}                 & \delta_{SO}~~~  & 2i\delta_{R} & 0 \\
0                           & -2i\delta_{R}~~~ & \delta_{SO}   & -i\delta_{0} \\
-2i\lambda_{R}              & 0~~~            & i\delta_{0}   & \delta_{SO}-2\lambda_{SO}\\
\end{array}
\right)
\end{equation}
First term in Eq.\ref{Eff} indicates a strain dependent constant energy shift which could be scaled out in the numerical calculations. Second term stands for strain modified intrinsic SOC. Third and forth term represent the generalized Rashba interaction. Note that the unstrained Rashba coupling has been given by $H_{\mathcal{R}}^{(0)}=\lambda^{(0)}_{R}(\tau \sigma_{x}s_{y}-\sigma_{y}s_{x})$, where $\lambda^{(0)}_{R}=\lambda_{R}(\epsilon=0)$. By making a comparison between the strained and unstrained Rashba interactions, it can be realized that unlike the symmetric form of the Rashba interaction in the unstrained sample, uniaxial strain deforms the Rashba interaction asymmetrically. This asymmetry characterizes by $\delta_R$ which measures the difference between the spin-flip hopping amplitudes of $x$ and $y$ directions. Uniaxial strain emerges a new term that characterizes by the coupling strength of $\delta_{0}$. This term corresponds to the intra-sublattice spin-flips ($|A p_z \uparrow \rangle \leftrightarrow|A p_z \downarrow\rangle$ or $|B p_z \uparrow \rangle \leftrightarrow|B p_z \downarrow\rangle$) transitions.

In the unstrained graphene (where $\delta_{R}$, $\delta_{0}$ and $\delta_{SO}$ are zero) it can be realized that the Rashba interaction is responsible for $|A p_z\uparrow\rangle \leftrightarrow|B p_z\downarrow\rangle$ and $|A p_z\downarrow\rangle \leftrightarrow|B p_z\uparrow\rangle$ transitions in $K$ and $K'$ Dirac points, respectively. However, one can figure out that the uniaxial strain makes both of these transitions possible at either of Dirac points. Meanwhile, Rashba coupling strength itself, exponentially increases by positive strains as can bee seen in Figs. \ref{Fig_param_x} and \ref{Fig_param_y}. 

\begin{figure*}
\resizebox{0.8\textwidth}{!}{\includegraphics[width=1.0\linewidth]{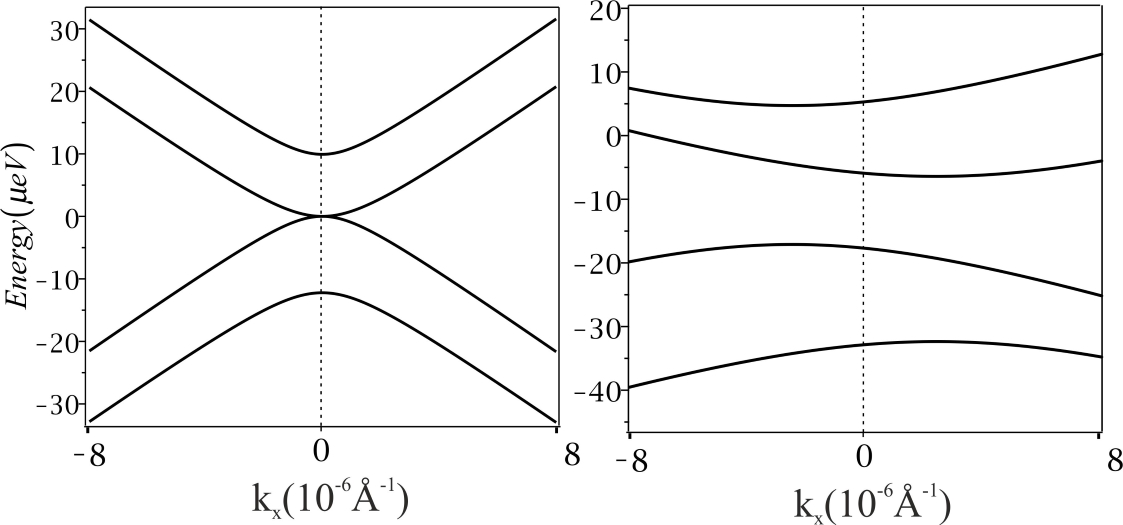}
}
\centering
\caption{Tight-binding energy spectrum of (a) unstrained and (b) +20\% zigzag strained graphene by taking into account the both intrinsic and external SOCs generated by a perpendicular electric filed of $E=1 V/nm$. 
	 \label{Fig_Spectrum} }
\end{figure*}

In addition, as can be seen in the Fig.\ref{Fig_Spectrum} strain induces indirect gap in the system. To formulate the k dependence of the effective Hamiltonian we have to add the spin-orbit coupling independent part of the effective Hamiltonian. If we use the approximate form of this part as $(\sigma .(\mathbbm{1}_{\sigma}+\epsilon-\beta \epsilon).\vec{k})\mathbbm{1}_{s}$ which proposed in Ref. \cite{mexic} energy bands near the Dirac points can be formulated as
\begin{equation}
\begin{array}{l}
E_{k}\simeq \lambda_{1} + \nu \sqrt{( \hbar \upsilon_{F}|\vec{q}| - \mu \delta_{0})^{2}+\lambda_{2}^{2}},
\end{array}
\end{equation}
$\upsilon_{F}$ is the Fermi velocity of the unstrained graphene, $\mu,\nu=\pm1$, $\vec{q}=(q_{x},q_{y})$ and
\begin{equation}
\begin{array}{l}

q_{x}=(1+\epsilon_{xx}-\beta \epsilon_{xx})k_{x}+(\epsilon_{xy}-\beta \epsilon_{xy})k_{y},
\\
q_{y}=(1+\epsilon_{yy}-\beta \epsilon_{yy})k_{y}+(\epsilon_{yx}-\beta \epsilon_{yx})k_{x},
\\
\lambda_{1}= \mu (\lambda_{R}-\delta_{R}) -\lambda_{SO} +\delta_{SO} ,
\\
\lambda_{2}=\lambda_{R}+\delta_{R} - \mu \lambda_{SO}.
\end{array}
\end{equation}
In the limit of $\epsilon=0$ mentioned band energy differs from the relation represented by Refs. \cite{FabianTopolgy50,deformation_17} by only a constant value. 

The linear energy spectrum of graphene around the Dirac points will not survive in the presence of SOC \cite{Fabian}. The energy spectrum of the system is reshaped into parabolic bands with a small energy gap as a result of SOC. By applying the transverse electric field, conduction and valence bands splits into four spin resolved bands.
 
\subsection{Dependence of the effective Hamiltonian on the electric field and strain }
Dependence of Rashba or intrinsic SOC strengths ($\lambda_{R}$ and $\lambda_{SO} $) on inter-atomic distance is studied by some authors \cite{Fabian,deformation_17}. This could describe the influence of homogeneous strains on spin-orbit couplings. In some other studies the intrinsic SOC as a function of strain have been investigated using first principle and symmetry based invariance approaches \cite{MoS_29,BAIHUA} .

In order to extract a general expression for the dependence of low-energy effective Hamiltonian on strain and external electric field it should be noticed that because the graphene is not intrinsically piezoelectric \cite{ACS_PizoE}, the influence of the electric field and strain can be considered independent of each other. In other words, perpendicular electric field does not induce strain and strain cannot induce perpendicular electric filed. Therefor, one can write the Hamiltonian parameters as $F(E)S(\epsilon)$ in which $F$ and $S$ represent the electric field and strain dependent parts of a given relation, respectively. Dependence of the effective Hamiltonian parameters on strain and electric field at the Dirac points has been summarized as fitted analytical relations in Tables \ref{tab:T_Parameters_x_new} and \ref{tab:T_Parameters_y_new} for zigzag and armchair strains, respectively.
\begin{table*}
	\caption{Dependence of the effective Hamiltonian parameters (in the unit of $\mu eV$) on zigzag strain and perpendicular electric field (in the unit of $1V/nm$). 	\label{tab:T_Parameters_x_new}}
	\begin{tabular}{lll}
		\hline
		\hline
		\\
		range of strain
		 ~~~~~~~~~~ & $\epsilon \leq 0$ & ~~~~~  $\epsilon \geq 0$ \\
		
		\hline
		\\
		$~\delta_{SO}(\epsilon,E)~$ & $~~0.43E^2(1-e^{-9.78\epsilon_{xx}})$~~ & $0.18E^2(1-e^{21.05\epsilon_{xx}})$ \\
		$~\delta_{R}(\epsilon,E)~$ &$~~E (1.3\epsilon_{xx}+8.7\epsilon_{xx}^2)$ &~~ $-0.09E(1-e^{16.98\epsilon_{xx}})$   \\
		\\
		\hline
		\\
		range of strain  & $ -0.2 \leq \epsilon \leq 0.2$ &  \\
		\hline
		\\
		$~\lambda_{SO}(\epsilon,E)~$ & $~~0.48+0.09e^{9.1\epsilon_{xx}}$ \\
		$~\delta_{0}(\epsilon,E)~$ & $~~2.43E(1-e^{4.4\epsilon_{xx}}) $  \\
		$~\lambda_{R}(\epsilon,E)~$ & $~~E(5.27+0.26e^{13.45\epsilon_{xx}})$  \\
		\normalsize
		\\
		
		\hline
		\hline
		
	\end{tabular}
\end{table*}

\begin{table*}
	\caption{ Dependence of the effective Hamiltonian parameters (in the unit of $\mu eV$) on armchair strain and perpendicular electric field (in the unit of $1V/nm$). \label{tab:T_Parameters_y_new}}
	\begin{tabular}{llll}
		\hline
		\hline
		\\
		range of strain ~~~~~~~~~~  & $\epsilon \leq 0$& ~~~~~ $\epsilon \geq 0$ \\
		\hline
		\\
		$\lambda_{R}(\epsilon,E)$ & $E(5.57+6.03 \epsilon_{yy}+39.27\epsilon_{yy})$&  ~~ $E(4.89+0.64 e^{ 7.33\epsilon_{yy}})$  \\
		$\delta_{SO}(\epsilon,E)$ & $0.23E^2(1-e^{-17.12\epsilon_{yy}})$& ~~~~~~ $0.34E^2(1-e^{12.70\epsilon_{yy}})$ \\
		$\delta_{R}(\epsilon,E)$ & $-0.12E(1-e^{-13.53\epsilon_{yy}})$& ~~~~~~ $E(-1.5 \epsilon_{yy}+9.9\epsilon_{yy}^2)$ \\
		\\
		\hline
		\\
		range of strain  & $ -0.2 \leq \epsilon \leq 0.2$ &  \\
		\hline
		\\
		$\lambda_{SO}(\epsilon,E)$ & $0.47+0.1e^{7.92\epsilon_{yy}}$  \\
		$\delta_{0}(\epsilon,E)$ & $-6.5E(1-e^{1.66\epsilon_{yy}})$  \\
		\\
		
		\hline
		\hline
		
	\end{tabular}
\end{table*}

\begin{figure*}
	\centering
	\includegraphics[width=0.7\linewidth]{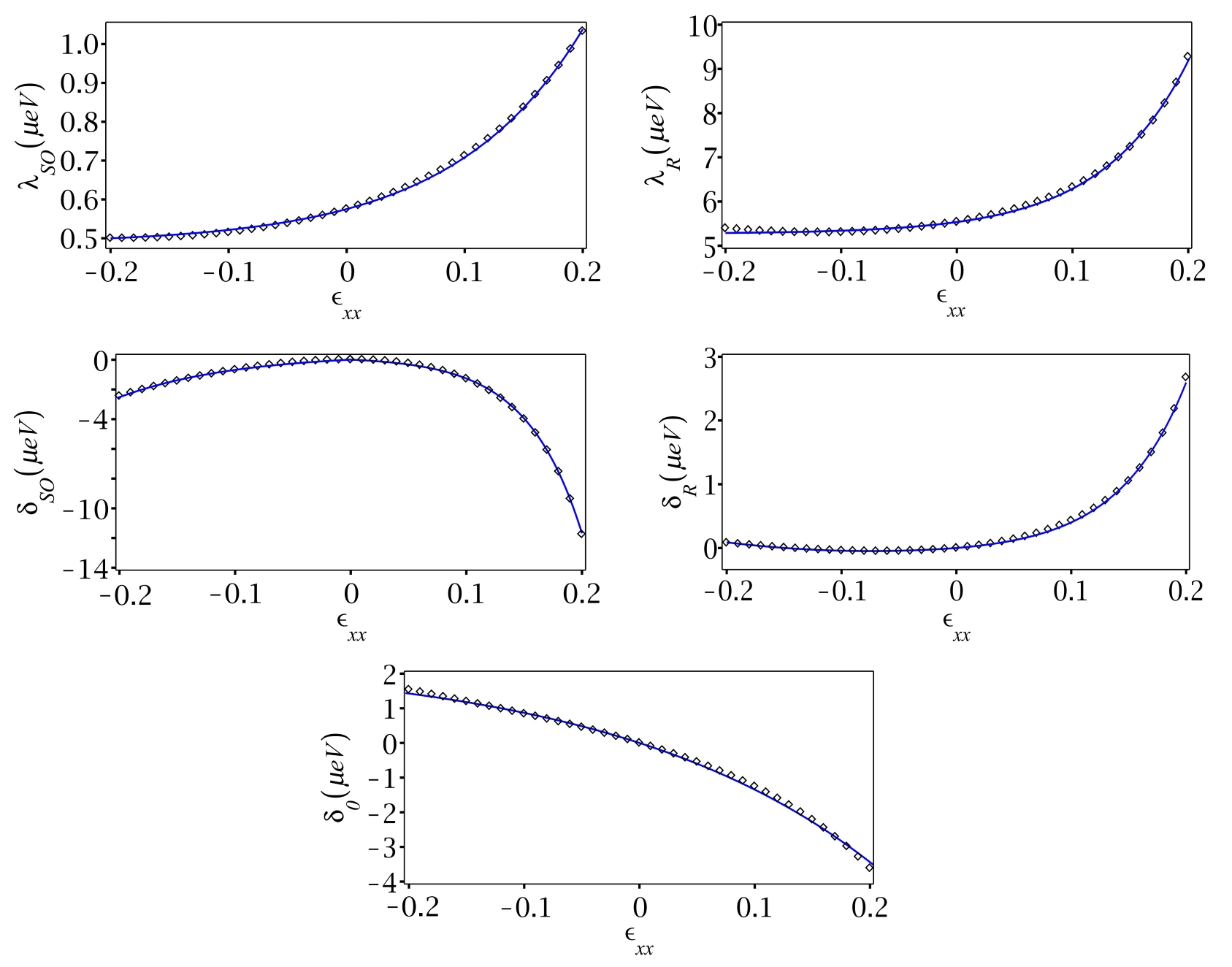}
	
	\caption{(Color online)Dependence of effective Hamiltonian parameters on zigzag strain. Solid lines refer to the fitted curves and diamonds show values derived by TB method \label{Fig_param_x}}
\end{figure*}

\begin{figure*}
	\centering
	\includegraphics[width=0.7 \linewidth]{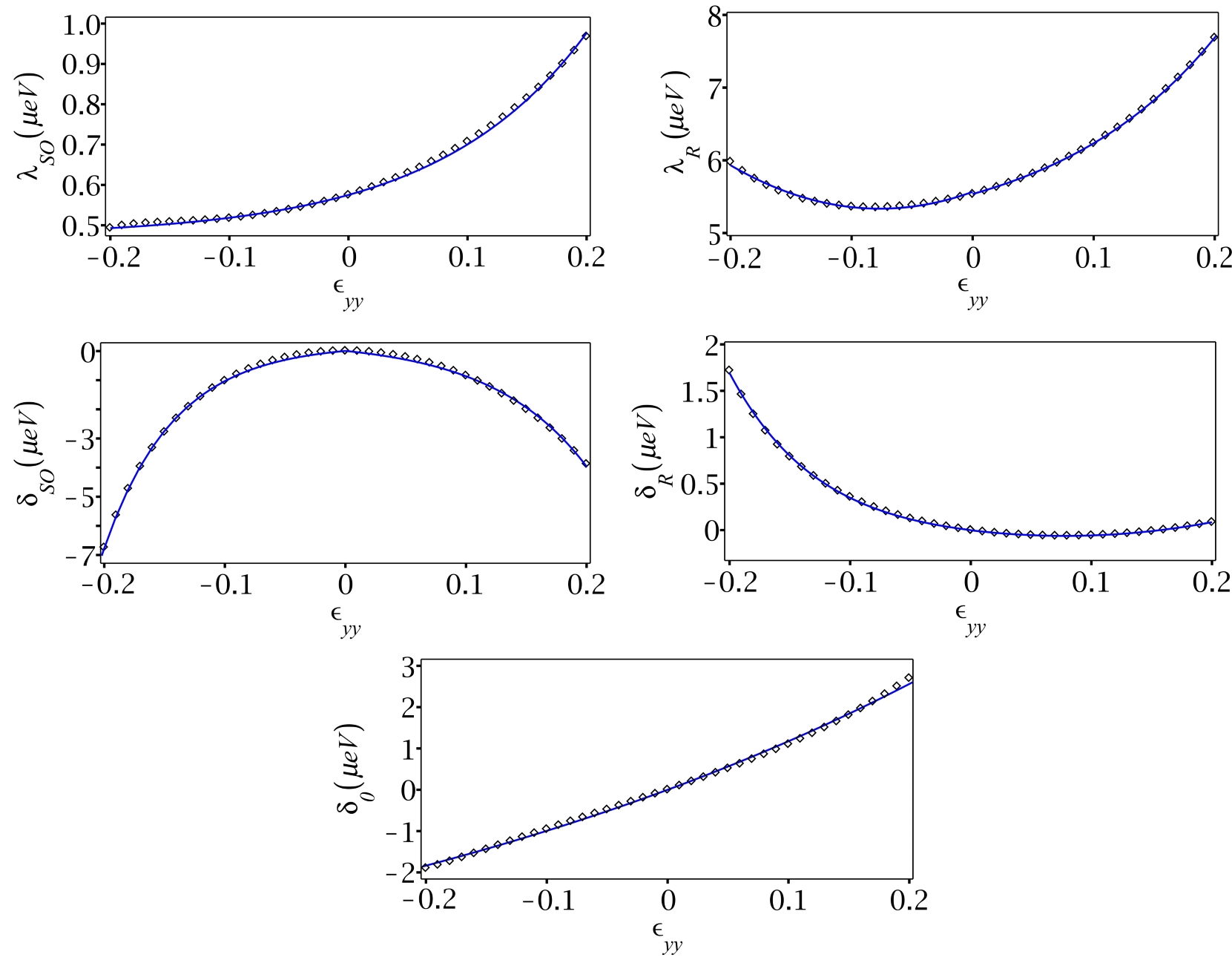}
	\centering
	
	\caption{(Color online)Dependence of effective Hamiltonian parameters on armchair strain. Solid lines refer to the fitted curves and diamonds show values derived by TB method \label{Fig_param_y}}
\end{figure*}

The dependence of effective Hamiltonian on strain has also been investigated by fixing the external electric field in typical available value of $E=1V/nm$ \cite{deformation_17,Fabian}. This Dependence can be analytically formulated by fitting exponential or quadratic functions to the numerical results as depicted in Fig \ref{Fig_param_x} and Fig  \ref{Fig_param_y}. Fitted functions has been characterized in Table \ref{tab:T_Parameters_x_new} and Table \ref{tab:T_Parameters_y_new}.
 
By increasing the tensile strain, $\lambda_{SO}$ exponentially increases but compressive strain cannot change the $\lambda_{SO}$ considerably. With a $+20$\% strain $\lambda_{SO}$ increases up to $80$\%. Meanwhile, $\lambda_{R}$ increases by increasing the tensile strain up to $68$\%. This parameter does not show considerable change for compressive zigzag strain (variation is less than $7$\%). By increasing the tensile zigzag strain, $\delta_{R}$ increases up to $3\mu eV$. In addition, $\delta_{R}$ increases by compressive armchair strain up to $1.8\mu eV$. However, as shown in Figs. \ref{Fig_param_x} and \ref{Fig_param_y} compressive zigzag and tensile armchair strains cannot cause considerable change in the $\delta_{R}$ value. In the strain range of $-20\% $ to $+20\%$ $\delta_{0}$ shows approximately linear change from $2\mu eV$ to $-3\mu eV$ for zigzag strain and from  $-2\mu eV$ to $+3\mu eV$ for armchair strain. Parameters $\lambda_{R}$, $\delta_{R}$ and $\delta_{0}$ are proportional to external electric field. This fact can be compared with linear dependence of Rashba parameter $\lambda_{R}^{0}$ on electric filed in the unstrained graphene as reported by other authors
\cite{Fabian,FabianTopolgy50,deformation_17,Macdonald,nanotube_10}.

\subsection{Intrinsic Gap at zero electric field ($E=0$)}
The dependence of intrinsic SOC induced gap on the strain in the absence of external electric field has been obtained in the present investigation. There are different reports about the estimated value of intrinsic SOC gap at Dirac points \cite{Fabian,Macdonald}. 
It should be noticed that the intrinsic gap is given by $\Delta_{SO}=2\lambda_{SO}$ when the Rashba coupling is zero ($E=0$) even in the strained graphene. Accordingly, same as $\lambda_{SO}$ intrinsic gap has an exponential dependence on the strain. As it can be seen in Figs.\ref{Fig_param_x} and \ref{Fig_param_y} for negative (Compressive) strains intrinsic band gap approximately remains constant. It can be inferred from the above relation that the intrinsic gap can be increased up to the $80\%$ by a strain of $+20\%$. In addition one can obtain that intrinsic gap in unstrained graphene is 1.14 $\mu eV$ same as previous results \cite{Macdonald,firstPrncpl9}.
\subsection{External gap in the presence of the vertical electric field ($E\neq0$)}
 As discussed before, in the presence of a transverse external electric field there will be a Rashba-type SOC in the graphene. In the absence of strain for $\lambda_{R} > \lambda_{SO}$, SOC induced energy gap falls to zero \cite{FabianTopolgy50,Macdonald,kane7}. However, results of the present study show that even in this condition energy gap could be induced by uniaxial strain.
 \\
 
 By analyzing the eigenvalues of effective Hamiltonian in the minimum of conduction band and maximum of valence band we can represent the relation between external energy gap and effective Hamiltonian parameters as
 \begin{equation}
 \label{gap_general}
  E_{gap}=
 \begin{cases} 
 2(\lambda_{SO} -\lambda_{R}-\delta_{R}) & \lambda_{R}\leq \lambda_{SO} -\delta_{R} \\
 2(\lambda_{R}+\delta_{R}-\lambda_{SO} ) & \lambda_{SO}-\delta_{R}\leq \lambda_{R}\leq \lambda_{SO}+\delta_{R} ,\\
 4\delta_{R} &  \lambda_{SO} +\delta_{R}\leq \lambda_{R} 
 \end{cases}
\end{equation} 
 where it can be noticed that if $\delta_{R}<0$ we have $E_{gap}=0$ and the gap is proportional to the external electric field for $\lambda_{SO} +\delta_{R}\leq \lambda_{R}$.
 
\begin{figure}[h]
 	\centering
 	\includegraphics[width=0.9\linewidth]{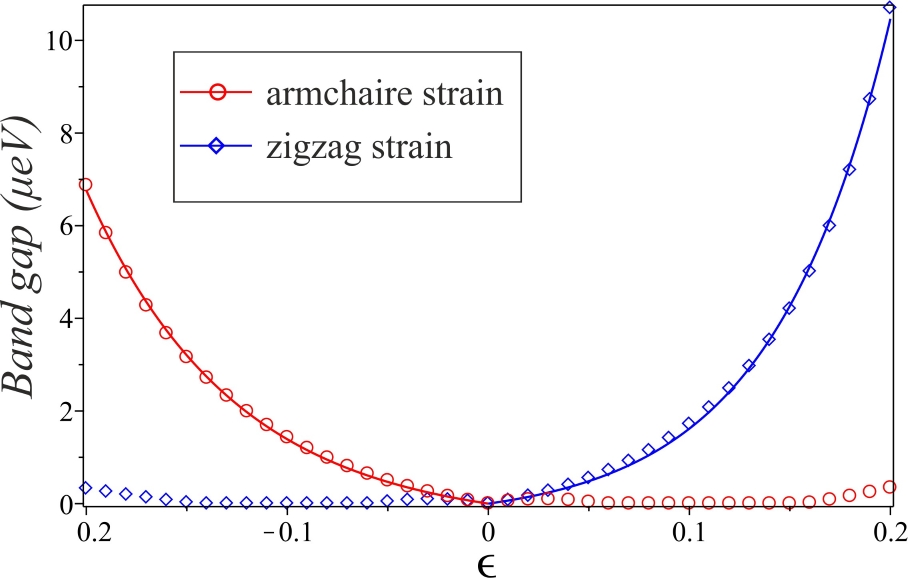}
 	\caption{(Color online) Energy gap of the strained graphene in the presence of typical electric field of $E=1~V/nm$ as a function of the strain strength in the zigzag (blue diamonds) and armchair (red circles) directions. Solid lines show the fitted curves.\label{indirect_gap}}
 \end{figure}

  Fig.\ref{indirect_gap} shows the calculated energy gap of the strained graphene in the presence of external transverse electric field with typical value of $1 V/nm$. It is obvious that a $+20\%$ zigzag strain can induce a splitting of order $10\mu eV$. We see that for zigzag strain, energy gap increases exponentially with increasing the amount of tensile strain while, for armchair strain the energy gap increases with increasing the amount of compressive strain. The band gap for the compressive zigzag strain and tensile armchair strain is negligible. By fitting the TB results to exponential function in the presence of $1 V/nm$ electric field it is possible to model the dependence of the band gap on strain in the specified strain range as
\begin{eqnarray}
 E_{gap}(\epsilon)= 0.36(  e^{16.98 \epsilon}-1)   (\mu eV), 
 \nonumber\\
 \text{ zigzag strain} ~~~,0\leqslant\epsilon\leqslant 0.2 
 \end{eqnarray}
 
 \begin{eqnarray} 
 E_{gap}(\epsilon)=0.48( e^{-13.53 \epsilon}-1)   (\mu eV),\nonumber\\  \text{armchair strain}, -0.2 \leqslant\epsilon\leqslant 0 .
 \end{eqnarray}
As it can be inferred these relations are in agreement with the general band gap expression given in Eq. [\ref{gap_general}] when $  \lambda_{SO} +\delta_{R}\leq \lambda_{R} $.

\section*{CONCLUSIONS}
We have proposed a generalized low-energy effective Hamiltonian for the intrinsic and Rashba spin-orbit couplings in uniaxially strained graphene. By taking into account the deformation effect on Slater-Koster parameters we have formulated the strain modified SOCs in the graphene. Results of the present study show that both intrinsic and Rashba SOCs could be effectively increased by the applied strain. We show that uniaxial strain introduces a deviation of the Rashba interaction from its symmetric form of unstrained system. This deviation could be characterized by a new parameter. We have shown that SOCs induced band splittings can be tuned by changing the magnitude and direction of the strain. Dependence of the effective Hamiltonian parameters on the external electric field has also been investigated in the present study. Besides, numerical results show that, it is possible to manipulate the form of energy dispersion in graphene.

The system response is meaningfully anisotropic for the uniaxial strains. This is due to the anisotropic nature of the uniaxial strain which has been imposed into the real space configuration and also orbitals mutual orientation. Accordingly, Hamiltonian characteristic parameters show different dependence on zigzag and armchair uniaxial strains. Uniaxial strains induce anisotropy both in the the real space atomic configuration and also inter atomic orbitals overlap. Because the orbital orientation in the deformed lattice is not merely determined by the inter-atomic distance, one should take into account this additional source of the anisotropy.

The SOC is responsible for small band gap of graphene and uniaxial strains could effectively change the magnitude of the SOC induced band gap of the system. Therefore, the band gap could be controlled by the strain. Besides, it was realized that zigzag and armchair strains give different functionality of band energy gap. This is due to the same anisotropy which has been discussed before. 

Another important point which should be addressed in the present investigation is that the uniaxial strain breaks the symmetry of the conduction and valence bands in which the strain induces an indirect band gap in the sample.
As mentioned before in unstrained graphene the Rashba interaction is responsible for 
$|A p_z\uparrow\rangle \leftrightarrow|B p_z\downarrow\rangle$  transitions in $K$ point and
$|A p_z\downarrow\rangle \leftrightarrow|B p_z\uparrow\rangle$ transitions in $K'$ Dirac point. It has been realized that the uniaxial strain deforms the symmetric dependence of the Rashba coupling so that the above valley resolved picture of spin and pseudo-spin transitions in the unstrained graphene has been completely destroyed. In the other words both type of the mentioned transitions are possible (however, with different transition amplitudes) at each of the Dirac points in strained graphene.

Because the functionality and strength of the spin-orbit couplings can be controlled by the amount and direction of the applied strain (regarding the importance of the strain engineering in the valley and pseudo-spin polarization), it can be expected that the strain may play the same important role in the field of spintronics as it plays in subject of valleytronics and pseudo-spintronics \cite{Morgenstern,jiang2013}. In addition, spin-polarization at the boundaries \cite{njp} of finite-width graphene nano-ribbon, in non-equilibrium regime could effectively control pseudo-spin polarization via the Rashba interaction that couples electron spin and pseudo-spin degrees of freedom. Meanwhile, it could be interesting to determine the possibility of spin and  pseudo-spin exchange by Rashba interaction, since the Rashba coupling strength can be modulated by external strain. 
\section*{Acknowledgment}
This research has been supported by Azarbaijan Shahid Madani university.
\section*{Author contribution statement}
Calculations have been performed by H. Rezaei. The paper has also be written by  H. Rezaei. A. Phirouznia supervised the
study and also revised the article. 
\begin{appendices}
\section{L\"{o}wdin transformation}
 For a block shaped Hamiltonian given by:

\begin{equation}
H=
\left(
\begin{array}{cc}
H_{0} & T \\
T^{\dag} & \Delta \\
\end{array}
\right),
\end{equation}
where $H_{0}$ and $\Delta$ are Hamiltonian matrix representation in different subspaces and $T$ represents the coupling of these two subspaces, $H$ can be reduced within the L\"{o}wdin method to an effective Hamiltonian of a subspace in which the $H_0$ has been presented. In this method we interested in $T$-coupling modified $H_{0}$ bands and it was assumed that the matrix elements of block $T$ are small relative to $\Delta$ eigenvalues. 

Consider a unitary matrix $S$ that transforms the Hamiltonian to a block-diagonal matrix $\tilde{H}$:
\begin{eqnarray}
\label{a2}
\tilde{H}&=&e^{-S} H e^{S}\nonumber\\ &\simeq& H+[H,S]+1/2[H,[H,S]],
\end{eqnarray}
where
\begin{equation}
S=
\left(
\begin{array}{cc}
0 & M \\
-M^{\dag} & 0 \\
\end{array}
\right),
\end{equation}
in which $M$ is a arbitrary Matrix.

Since transformed Matrix $\tilde{H}$ must be block-diagonal, $M$ must be determined by this equation:
\begin{equation}
T+H_{0}M-M\Delta+MT^{\dag}M=0 .
\end{equation}
If we just keep second order of $\Delta^{-1}$ matrix $M$ will be :

\begin{equation}
M \simeq T\Delta^{-1}+H_{0}T\Delta^{-2} .
\end{equation}
Now by neglecting the higher-order terms in $\Delta^{-1}$ and by using the Eq. \ref{a2} for $\tilde{H}$, first block of the effective Hamiltonian is given by:

\begin{equation}
H_{eff}\approx H_{0}-T \Delta^{-1} T^{\dag},
\end{equation}
which is the effective Hamiltonian in $H_0$ related subspace.
\end{appendices}
\bibliographystyle{spphys}
\bibliography{new}
%
\end{document}